\documentclass[11pt]{article}
\usepackage{comment}
\usepackage[margin=2.5cm]{geometry}
\usepackage[toc,page]{appendix}
\usepackage{amsmath,amsfonts,breqn,comment,cancel,mathtools,amssymb,extarrows,mathtools,graphicx,subfigure,setspace,bbold,physics}
\usepackage{xr-hyper} 
\usepackage{cite}
\usepackage{slashed}
\usepackage[dvipsnames]{xcolor}
\usepackage{hyperref}
\hypersetup{colorlinks=true, linkcolor=blue, citecolor=magenta, linktoc=page}
\usepackage{url}
\usepackage{tikz}
\usepackage{color}
\usepackage{arydshln,leftidx,mathtools}
\usetikzlibrary{decorations.pathreplacing}
\numberwithin{equation}{section}
\newcommand{\be}{\begin{equation}}
	\newcommand{\bea}{\begin{eqnarray}}
		\newcommand{\eea}{\end{eqnarray}}
	\newcommand{\ba}{\begin{align}}
		\newcommand{\ea}{\end{align}}
	\newcommand{\ee}{\end{equation}}

\usepackage{chngcntr}
\begin{document}
	\onehalfspacing
\begin{titlepage}
	\thispagestyle{empty}

%
	
	\vspace{.4cm}
	\begin{center}
		\noindent{\Large \textbf{Capacity of entanglement for scalar fields in squeezed states}}\\
		
		\vspace*{15mm}
		\vspace*{1mm}
		\vspace*{1mm}
		{M. Reza Mohammadi Mozaffar}
		
		\vspace*{1cm}
		
		{\it Department of Physics, University of Guilan,
			P.O. Box 41335-1914, Rasht, Iran
		}
		
		\vspace*{0.5cm}
		{E-mail: {\tt mmohammadi@guilan.ac.ir}}
		
		\vspace*{1cm}
	\end{center}
	
	\begin{abstract}

We study various aspects of capacity of entanglement in the squeezed states of a scalar field theory. This quantity is a quantum informational counterpart of heat capacity and characterizes the width of the eigenvalue spectrum of the reduced density matrix. In particular, we carefully examine the dependence of capacity of entanglement and its universal terms on the squeezing parameter in the specific regimes of the parameter space. Remarkably, we find that the capacity of entanglement obeys a volume law in the large squeezing limit. We discuss how these results are consistent with the behavior of other entanglement measures including entanglement and Renyi entropies. We also comment
on the existence of consistent holographic duals for a family of Gaussian states  with generic squeezing parameter based on the ratio of entanglement entropy and the capacity of entanglement.
\end{abstract}

\end{titlepage}
\newpage

\tableofcontents
\noindent
\hrulefill


\section{Introduction}\label{intro}

In recent years, surprising new connections have been developing between quantum information theory, quantum many-body systems and quantum gravity. In particular, understanding the entanglement structure of quantum systems in a pure or mixed state has become an active area of research, \textit{e.g.}, see \cite{Eisert:2008ur,Casini:2009sr,Calabrese:2009qy,Laflorencie:2015eck,Nishioka:2018khk,Casini:2022rlv} for reviews. Also in the context of gauge/gravity correspondence, fascinating connections have been developing between quantum information and quantum gravity including the holographic entanglement entropy proposals and the notion of geometry from entanglement, \textit{e.g.}, see \cite{Rangamani:2016dms} and references therein. Moreover, in order to quantify entanglement and quantum correlations several measures has been studied so far including entanglement and Renyi entropies. Indeed, the entanglement entropy is the unique measure which assesses the amount of quantum entanglement between two subsystems for a given pure state $|\psi\rangle$. In this case, assuming that $\mathcal{H}_{\rm tot.}=\mathcal{H}_{A}\otimes \mathcal{H}_{\bar{A}}$, entanglement entropy can be written in terms of von Nuemann entropy as
\begin{eqnarray}\label{EE}
S_{E}=-{\rm Tr}_A\left(\rho_A \log \rho_A\right),
\end{eqnarray}
where $\rho_A$ is the reduced density matrix defined as $\rho_A={\rm Tr}_{\bar{A}}\left(|\psi\rangle\langle\psi|\right)$. Further, the Renyi entropy is a one-parameter generalization of entanglement entropy which is given by
\begin{eqnarray}\label{renyi}
S_{n}=\frac{1}{1-n}\log {\rm Tr}\rho_A^n,
\end{eqnarray}
where $n$ is a positive integer. It is then easy to show that entanglement entropy follows by analytic continuation of the Renyi entropy down to $n=1$, \textit{i.e.}, $S_E=\lim_{n\rightarrow 1}S_n$. Further, other interesting cases to consider are the large and small $n$ limits, which yield
\begin{eqnarray}\label{renyilargen}
S_{\infty}\equiv \lim_{n\rightarrow \infty} S_n=- \log  \lambda_{\rm max},\hspace*{1.5cm}S_{0}\equiv \lim_{n\rightarrow 0} S_n=\log  \mathcal{N},
\end{eqnarray}
where $\lambda_{\rm max}$ is the largest eigenvalue of the reduced density matrix and $\mathcal{N}$ denotes the number of nonvanishing eigenvalues of $\rho_A$. Besides the already mentioned case of the entanglement and Renyi entropies, there are many attempts to construct new information theoretic measures for studying the entanglement structure in more general setting, \textit{e.g.}, mutual information \cite{Casini:2008wt}, logarithmic negativity \cite{Vidal,Plenio:2005cwa} and entanglement of purification \cite{Terhal:2002}. However, in this paper, we focus on another measure that has recently entered this discussion which is called the capacity of entanglement \cite{Yao:2010woi}
\begin{eqnarray}\label{capa}
C_E\equiv \lim_{n\rightarrow 1} C_n=\lim_{n\rightarrow 1} n^2\frac{\partial^2}{\partial n^2}\left((1-n)S_n\right),
\end{eqnarray}
where $C_n$ is the $n$-th capacity of entanglement.
Indeed, considering $n$ as the inverse temperature, the above definition is similar to the corresponding relation between the heat capacity and thermal entropy. Further, using the definition of the modular Hamiltonian, \textit{i.e.}, $H_A=-\log \rho_A$, it is straightforward to show that the entanglement entropy and capacity of entanglement are the expectation value and variance (the second cumulant) of $H_A$ respectively 
\begin{eqnarray}\label{secapa}
&&S_E={\rm Tr}\;\rho_A H_A=\langle H_A \rangle,\nonumber\\
&&C_E={\rm Tr}\;\rho_A H_A^2-\left({\rm Tr}\;\rho_A H_A\right)^2=\langle H_A^2 \rangle-\langle H_A \rangle^2.
\end{eqnarray}
Thus the capacity of entanglement characterizes the width of the eigenvalue spectrum of the reduced density
matrix such that for a maximally entangled state, \textit{i.e.}, $\rho_A\propto I$, it vanishes. Indeed, in this case the Renyi entropies are independent of $n$ and we have a flat entanglement spectra. Recently, there have been many attempts to investigate various properties of this quantity in different setups both in the field theory and holography which have led to a remarkably rich and varied range
of new insights, \textit{e.g.}, \cite{Nakaguchi:2016zqi,Nakagawa:2017wis,DeBoer:2018kvc,Okuyama:2021ylc,Nandy:2021hmk,Bhattacharjee:2021jff,Wei:2022bed,Shrimali:2022bvt,Arias:2023kni,Andrzejewski:2023dja,Ren:2024qmx,Banks:2024cqo,Shrimali:2024nbc}. In particular, an interesting observation in \cite{DeBoer:2018kvc} was that systems where are all entanglement is carried by EPR pairs have zero capacity of entanglement. Thus, whenever we find that $C_E\sim S_E$, \textit{e.g.}, $(1+1)$-dimensional CFTs at equilibrium, EPR pairs are not a very good approximation of the quantum state and randomly entangled pairs of qubits give a better description. Moreover, as proposed in this reference, restricting to field theories with
holographic duals without higher derivative terms, the ratio $\frac{C_E}{S_E}$ turns out to be exactly equal to one. An interesting question is whether this result also hold for more general QFTs without a holographic dual.

Clearly, the capacity of entanglement depends on the choice of the original state of the system. The main aim of this article is to further investigate the state dependence of the capacity of entanglement. We study this dependence by considering a general class of Gaussian states which are solutions to the time dependent Schr\"{o}dinger equation, the so-called squeezed states. As we will see the squeezed states are less classical than the vacuum state (which is a special case of a coherent state) and thus study the entanglement structure in a squeezed state may help us to gain a better insight into the quantum features of the system in question. In the following we will focus on a scalar field theory in $(1+1)$-dimensions to address this problem. Indeed, in order to avoid the ultraviolet divergences in the continuum limit, we should regulate the theory by placing it on a one dimensional spatial lattice. Thus the model reduces to an infinite chain of quantum harmonic oscillators. Let us add that the entanglement entropy for this setup in squeezed states has been previously studied in \cite{Katsinis:2023hqn}. These authors proposed an extension of the method introduced in \cite{Bombelli:1986rw,Srednicki:1993im} to more general cases including squeezed states. Interestingly, using a systematic method they show that for a free scalar field theory in $(1+1)$-dimensions the entanglement entropy obeys the volume law rather than the area law in large squeezing limit. This is in contrast with the behavior of entanglement entropy in typical ground states which exhibits area law scaling where in $(1+1)$-dimensions is replaced with a logarithmic scaling. Further, their result is in agreement with the Page's argument where for a typical pure quantum state of a joint system, the smaller subsystem is almost maximally mixed, showing little sign that the total system is pure \cite{Page:1993df}. Interestingly, the Page curve followed by the entropy of Hawking radiation \cite{Page:1993wv}. In this context, an especially interesting question concerns how the scaling of the capacity of entanglement alters under a non-trivial squeezing when the total system is in a random squeezed state.

Additionally, in \cite{Kawabata:2021hac,Kawabata:2021vyo}, using the holographic proposals, it was shown that the capacity of entanglement is a useful probe of the Hawking radiation during the black hole evaporation process. Interestingly enough, the authors argued that this measure shows a discontinuity or a peak at the Page time while the entanglement entropy varies smoothly. In this sense, one can consider $C_E$ as a convenient probe which can characterizes the order of a phase of the quantum systems. Indeed a primary motivation for this work came from these efforts to study the behavior of different entanglement measures including the capacity of entanglement in random pure states and to compare our results to those obtained in the context of holography. Let us mention that this comparison involves two very different models, that is a free scalar theory with a single degree of freedom versus a holographic quantum field theory which is strongly coupled and has a large number of degrees of freedom.

The remainder of our paper is organized as follows: In section \ref{sec:setup}, we give the general framework
in which we are working, establishing our notation and the general form of the squeezed states and harmonic models in question. Section \ref{sec:prelim} contains a summary about the behaviors of entanglement measures in ground state of a scalar theory. We review old results for the entanglement entropy and also find new ones for the case of capacity of entanglement. To get a better understanding of the results, we will also compare the behavior of different measures in specific scaling regimes. In section \ref{sec:harmonicsqueez}, we extend our studies to the case of squeezed states. Specifically, we present a combination of numerical and analytic results on the scaling of different entanglement measures in the large squeezing limit. Next, we return to the field theory problem by generalizing these results to the continuum limit in section \ref{sec:QFTsqueez}. Given the entanglement measures for the scalar field theory, we then ask how our results compare to holographic ones. Finally, we close in section \ref{sec:diss} with a brief discussion of our results and directions for future work.

\section{Set-up}\label{sec:setup}

We start by considering a simple quantum harmonic oscillator with the following Hamiltonian\footnote{For simplicity of notation in what follows, we will use hat symbol for the quantum operators only when there is a danger of confusion. Also we will set $\hbar=1$.}
\begin{equation}\label{hamilosci}
H=\frac{p^2}{2m}+\frac{1}{2}m\omega^2x^2.
\end{equation}
Defining the annihilation and creation operators 
\begin{equation}\label{aadagger}
a=\frac{1}{\sqrt{2m\omega}}(m\omega x+ip),\hspace*{2cm} a^\dagger=\frac{1}{\sqrt{2m\omega}}(m\omega x-ip),
\end{equation}
we have $H=\omega(a^\dagger a+\frac{1}{2})$.
In this case a class of Gaussian solutions to the time dependent Schr\"{o}dinger equation which consists of the so-called coherent states are given by
\begin{equation}\label{cohstate}
\psi(x, t)=\left(\frac{m\omega}{\pi}\right)^{1/4}\exp\left(-\frac{m\omega}{2}(x-x_0(t))^2+ip_0(t)(x-x_0(t))-i\phi_c(t)\right),
\end{equation}
where
\begin{equation}\label{cohstate1}
x_0(t)=X_0{\rm Re}\left(e^{i\omega(t-t_0)}\right),\hspace*{1.2cm}p_0(t)=m\dot{x}_0(t),\hspace*{1.2cm}\phi_c(t)=\frac{-x_0(t)p_0(t)}{2}+\frac{\omega(t-t_0)}{2}+\phi_0.
\end{equation}
These states minimize the uncertainty relation with uncertainty equally distributed between position and momentum such that
\begin{equation}\label{cohstate2}
\Delta x=\frac{\Delta p}{m\omega}=\frac{1}{\sqrt{2m\omega}},\hspace{2cm}\Delta x\,\Delta p=\frac{1}{2}.
\end{equation}
Indeed, they have the same minimal uncertainty value as found for the vacuum state. Further, the mean values of position and momentum operators in coherent state follow a classical orbit, \textit{e.g.}, $\langle \ddot{x}\rangle+\omega^2 \langle x\rangle=0$, and hence these states correspond to the most classical states of the harmonic oscillator. It is also easy to show that any eigenstate of the annihilation operator, \textit{i.e.}, $a|\alpha\rangle=\alpha|\alpha\rangle$, where $\alpha$ is a complex parameter, represents a coherent state. Clearly, these states include the vacuum state as a special case with $\alpha=0$. Moreover, defining a unitary displacement operator $D(\alpha)=e^{\alpha a^\dagger-\alpha^* a}$, the coherent states can be generated from the vacuum, \textit{i.e.,} $|\alpha\rangle=D(\alpha)|0\rangle$.

Let us now turn to another class of Gaussian solutions to the time dependent Schr\"{o}dinger equation which are more general, the so-called squeezed states. The corresponding wave function reads
\begin{equation}\label{squstate}
\psi(x, t)=\left(\frac{m{\Re}(w(t))}{\pi}\right)^{1/4}\exp\left(-\frac{mw(t)}{2}(x-x_0(t))^2+ip_0(t)(x-x_0(t))-i\phi_s(t)\right),
\end{equation}
where\footnote{We set the origin of time coordinate to be 0.}
\begin{eqnarray}\label{squstate1}
w(t)=\omega\frac{1-i\sinh z\;\cos 2\omega t}{\cosh z+\sinh z\;\sin 2\omega t},\hspace*{1cm}
\phi_s(t)=\phi_c(t)-\frac{\omega t}{2}+\frac{1}{2}\tan^{-1}\frac{\tanh\frac{z}{2}+\tan \omega t}{1+\tanh\frac{z}{2}\tan \omega t}.
\end{eqnarray}

In the above expressions $z$ is the squeezing parameter such that for $z=0$, we recover the coherent states. The key feature of a squeezed state in the harmonic potential is that although its profile is still Gaussian, its width is different from the vacuum state. Further, the corresponding uncertainty relation is easily found to be
\begin{equation}\label{squstate2}
\Delta x\,\Delta p=\frac{1}{2}\sqrt{1+\sinh^2z\;\cos^22\omega t},
\end{equation}
which shows that the squeezed states are not minimal uncertainty states at all times. Indeed, the time slices where the above relation becomes minimal are easily found to be
$t_{\rm min}=(2k+1)\frac{T}{8}$ where $T=\frac{2\pi}{\omega}$ and $k$ is an integer.
It is straightforward to show that when the squeezed state is a minimal uncertainty state, we have 
\begin{equation}
(\Delta x)_{\rm min}=\frac{1}{\sqrt{2m\omega e^{z}}},\hspace*{1.5cm}(\Delta p)_{\rm min}=\sqrt{\frac{m\omega e^{z}}{2}},
\end{equation}
which shows the possibility of arbitrary compression of the position uncertainty at the expense of appropriate fluctuation in the momentum variable and vice-versa. Interestingly, defining a unitary squeezing operator $S(\xi)=e^{\frac{1}{2}(\xi a^{\dagger2}-\xi^* a^2)}$, the squeezed states can be generated from the vacuum, \textit{i.e.,} $|\xi\rangle=S(\xi)|0\rangle$ where $\xi$ denotes a complex parameter.\footnote{Note that this definition gives a squeezed vacuum state. In addition one can obtain a squeezed coherent state by acting the squeezing operator on a coherent state, \textit{i.e.}, $|\xi, \alpha\rangle=S(\xi)|\alpha\rangle=S(\xi)D(\alpha)|0\rangle$. For reviews on the main properties of coherent and squeezed states see \textit{e.g.} Refs. \cite{Gerry,Rosas-Ortiz:2018jum} and references therein. }

In the next sections much of our discussion will focus on the different aspects of entanglement measures when the corresponding state is given by eq. \eqref{squstate} to examine how the squeezing parameter can affect different quantities. To do so, we consider a free massive scalar field in $(1+1)$-dimensions with Hamiltonian
\begin{eqnarray}\label{scalarhamil}
H=\frac{1}{2}\int dx\left(\pi^2(x)+\left(\partial_x\phi(x)\right)^2+m^2\phi^2(x)\right).
\end{eqnarray}
In order to circumvent the divergences so as to obtain finite results for the measures, we should regulate the above model by placing it on a lattice, which reduces the system to an infinite chain of coupled harmonic oscillators. Indeed, the general Hamiltonian for a system of $N$ coupled (one-dimensional) harmonic oscillators can be written as
\begin{eqnarray}\label{Nhamil}
H=\frac{1}{2}\sum_{i=1}^N\textbf{p}^T.\textbf{p}+\frac{1}{2}\sum_{i, j=1}^N \textbf{x}^T.K.\textbf{x},
\end{eqnarray}
where 
\begin{eqnarray}\label{xp}
\textbf{x}^T=\left(\begin{matrix}
x_1, \cdots, x_N
\end{matrix}\right),\hspace*{2cm}\textbf{p}^T=\left(\begin{matrix}
p_1, \cdots, p_N
\end{matrix}\right),
\end{eqnarray}
and $K$ is a real symmetric matrix. Now, this suggests that we begin with an even simpler warm-up problem, namely, the case of two coupled harmonic oscillators
\begin{eqnarray}\label{2oscihamil}
H=\frac{p_1^2}{2}+\frac{p_2^2}{2}+\frac{k_0}{2}(x_1^2+x_2^2)+\frac{k_1}{2}(x_1-x_2)^2,
\end{eqnarray}
where
$k_1$
determines the strength of the coupling between the two oscillators such that weak and strong coupling regimes correspond to $k_1\ll k_0$ and $k_1\gg k_0$ limits respectively. We see that this simple set-up maintains some interesting features of our original problem which helps us to better investigate the continuum scalar theory. Let us recall that related investigations attempting to better understand the entanglement structure in general Gaussian states have also appeared in \cite{Adesso,Bianchi:2015fra,Bianchi:2021lnp}.

\section{Preliminaries: $C_E$ for ground state} \label{sec:prelim}

As a first step toward understanding different entanglement measures including capacity of entanglement in squeezed states, we would like to study the same quantities in a simpler set-up where the state is vacuum. We begin with the case of two coupled harmonic oscillators and then, having built up some intuition, we return to the more involved problem
by generalizing these results to a lattice of coupled oscillators.

\subsection{Two coupled harmonic oscillators}\label{sec:prelimtwoosci}

In this case the corresponding Hamiltonian is given by eq. \eqref{2oscihamil}. Indeed, the entanglement entropy in this model was studied in 
\cite{Bombelli:1986rw,Srednicki:1993im}. Here we would like to apply the techniques developed in these references
to examine the behavior of Renyi entropy and capacity of entanglement. To do so, one simply rewrites the Hamiltonian in terms of the canonical coordinates,
\begin{eqnarray}\label{2oscihamilnormalmodes}
H=\frac{1}{2}\left(p_+^2+\omega_+^2x_+^2+p_-^2+\omega_-^2x_-^2\right),
\end{eqnarray}
where
\begin{eqnarray}\label{xpm}
x_\pm=\frac{x_1\pm x_2}{\sqrt{2}}, \hspace*{1cm}\omega_+^2=k_0,\hspace*{1cm}\omega_-^2=\omega_+^2+2k_1.
\end{eqnarray}
Now we have two decoupled harmonic oscillators, and thus the vacuum state can be written as the product of the vacuum state wave functions for the two individual oscillators as follows
\begin{eqnarray}
\psi_0(x_+, x_-)=\left(\frac{\omega_+\omega_-}{\pi^2}\right)^{1/4}\exp\left(-\frac{\omega_+ x_+^2+\omega_- x_-^2}{2}\right).
\end{eqnarray}
The above expression can be rewritten in terms of the physical coordinates of the two oscillators 
and then the corresponding reduced density matrix can be found by integrating out the $x_1$ coordinate, \textit{i.e.,} $\rho_2(x_2; x'_2)=\int dx_1\psi_0(x_1, x_2)\psi_0^*(x_1, x'_2)$, which yields
\begin{eqnarray}\label{rho2}
\rho_2(x_2; x'_2)=\left(\frac{\gamma-\beta}{\pi}\right)^{1/2}\exp\left(-\frac{\gamma}{2}\left(x_2^2+{x'}_2^2\right)+\beta {x}_2{x'}_2\right),
\end{eqnarray}
where
\begin{eqnarray}
\beta=\frac{\left(\omega_+-\omega_-\right)^2}{4\left(\omega_++\omega_-\right)},\hspace*{2cm}\gamma=\frac{\left(\omega_++\omega_-\right)^2+4\omega_+\omega_-}{4\left(\omega_++\omega_-\right)}.
\end{eqnarray}
Moreover, solving the eigenvalue problem for the reduced density matrix, \textit{i.e.,} 
\begin{eqnarray}
\int dy\; \rho_2(x; y)f_k(y)=p_kf_k(x),
\end{eqnarray}
one finds an infinite tower of eigenvalues as follows
\begin{eqnarray}\label{pkxi}
p_k=(1-\xi)\xi^k,\hspace{2cm}\xi=\frac{\beta}{\gamma+\alpha}=\left(\frac{\sqrt{\omega_+}-\sqrt{\omega_-}}{\sqrt{\omega_+}+\sqrt{\omega_-}}\right)^2,\hspace{2cm}k=0,\cdots, \infty,
\end{eqnarray}
where $\alpha=\sqrt{\gamma^2-\beta^2}$. Note that based on the above result the weak and strong coupling regimes correspond to $\xi\rightarrow 0$ and $\xi\rightarrow 1$ limits respectively. Now using eq. \eqref{renyi}, we can evaluate the Renyi entropy which yields
\begin{eqnarray}\label{2oscirenyi}
S_n=\frac{1}{1-n}\log \sum_{k=0}^\infty p_k^n=\frac{1}{1-n}\left(n\log(1-\xi)-\log(1-\xi^n)\right).
\end{eqnarray}
Combining the above result with eqs. \eqref{EE} and \eqref{capa}, we obtain entanglement entropy and capacity of entanglement
\begin{eqnarray}\label{2osciSECE}
S_E&=&\frac{\xi}{\xi-1}\log\xi-\log(1-\xi),\nonumber\\
C_E&=&\xi\left(\frac{\log\xi}{1-\xi}\right)^2.
\end{eqnarray}
Further the $n$-th capacity of entanglement is obtained by the replacement $\xi\rightarrow \xi^n$ in $C_E$. Remarkably, in the weak coupling regime a perturbative expansion yields
\begin{eqnarray}
S_E=-\xi\log\xi+\xi+\cdots,\hspace*{2cm} C_E=\xi\left(\log\xi\right)^2+2\xi^2\left(\log\xi\right)^2+\cdots,
\end{eqnarray}
which shows that in this limit both quantities vanish. This is consistent with the idea that as $k_1/k_0$ decreases, the reduced density matrix becomes more and more separable. On the other hand, in the strong coupling limit we obtain
\begin{eqnarray}\label{secestrong}
S_E=-\log(1-\xi)+1+\cdots\hspace*{2cm} C_E=1-\frac{(\xi-1)^2}{12}+\cdots.
\end{eqnarray}
Interestingly, we see that in this limit $C_E/S_E\ll 1$ which is consistent with the idea that the reduced density matrix becomes more and more maximally mixed as one increases the coupling.\footnote{Note that as we have already mentioned, for a maximally mixed state $C_E$ exactly vanishes. Although, here we follow the terminology of \cite{DeBoer:2018kvc} where the strength of the ratio between the capacity and entropy indicates how much a given reduced density matrix is close to a maximal distribution. Similar terminology used in \cite{Page:1993df} where the author encountered ``almost'' maximally mixed states. We would like to thank the anonymous referee for a useful comment
on this point.} 

\begin{figure}[h!]
	\begin{center}
\includegraphics[scale=0.59]{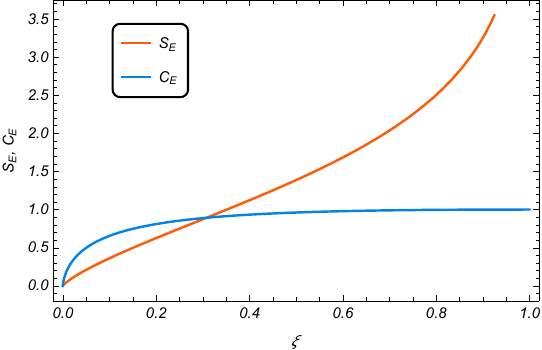}
\includegraphics[scale=0.59]{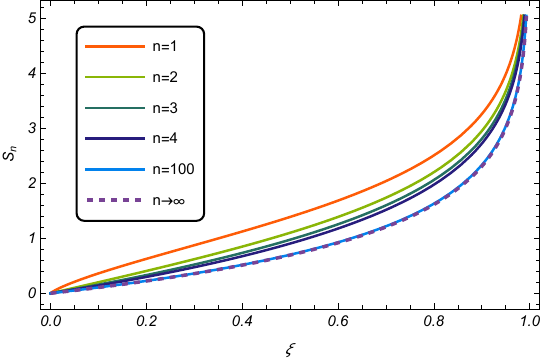}
\includegraphics[scale=0.59]{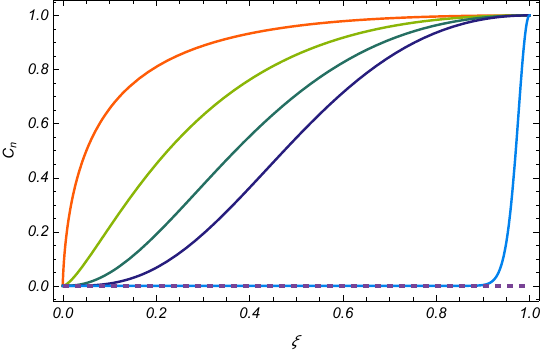}
  	\end{center}
	\caption{$S_E$ and $C_E$ (left), $S_n$ (middle) and $C_n$ (right) as functions of the coupling between the two harmonic oscillators. The dashed lines indicate the large $n$ limit which are consistent with eq. \eqref{largen}.}
	\label{fig:twoosci}
\end{figure}

\begin{figure}[h!]
	\begin{center}
\includegraphics[scale=0.87]{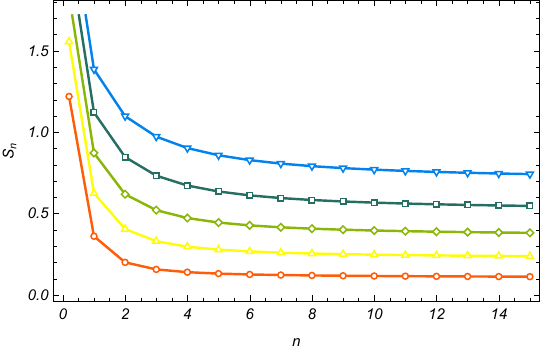}
\hspace*{0.4cm}
\includegraphics[scale=0.87]{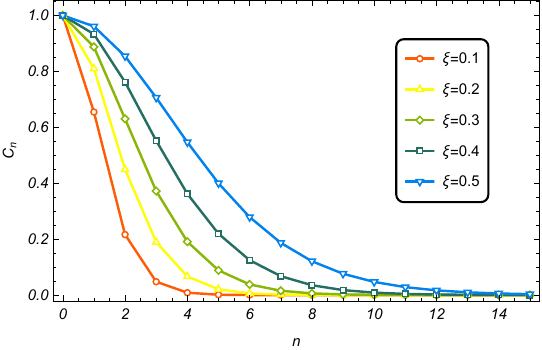}
  	\end{center}
	\caption{$S_n$ (left) and $C_n$ (right) as functions of the Renyi index for several values of the coupling between the two harmonic oscillators.}
	\label{fig:twoosci2}
\end{figure}

In figures \ref{fig:twoosci} and \ref{fig:twoosci2} we summarize the numerical results for different entanglement measures. Figure \ref{fig:twoosci} presents various quantities as functions of the coupling between the two harmonic oscillators. The left panel presents the dependence of entanglement entropy and capacity of entanglement on $\xi$. We note a number of key features: First, both these measures start at the same value (which is equal to zero corresponds to a separable state) and then increase as one increases the coupling. Second, although the entanglement entropy diverges in the strong coupling limit, the capacity of entanglement saturates to unity in agreement with eq. \eqref{secestrong}. Interestingly, for a specific value of the coupling these measures coincide, \textit{i.e., }$C_E(\xi_*)=S_E(\xi_*)$ where $\xi_*\sim 0.31$.  The middle panel demonstrates the Renyi entropy as a function of the coupling for several values of $n$. Although $S_n$ is a decreasing function of the Renyi index, it increases with the coupling as expected. In the right panel we show the $n$-th capacity of entanglement for the same values of the parameters. We see that $C_n$ decreases with $n$ and also saturates from below to unity.  

We present the $n$-dependence of $S_n$ and $C_n$ for several values of the coupling in figure \ref{fig:twoosci2}. Based on these plots, we see that the qualitative dependence of these measures on the coupling is similar to $n=1$ case. Also both measures decrease as we increase the Renyi index. Using eqs. \eqref{2oscirenyi} and \eqref{2osciSECE} it is easy to find the large $n$ scaling of these measures as follows 	
\begin{eqnarray}\label{largen}
S_{n\rightarrow\infty}\sim -\log(1-\xi),\hspace*{2cm}C_{n\rightarrow\infty}\sim \xi^n n^2(\log\xi)^2,
\end{eqnarray}
which shows that for any $0<\xi<1$, $C_{n\rightarrow\infty}$ vanishes. Indeed, our numerical results approach the above expressions in this limit as depicted in figure \ref{fig:twoosci}.
Finally, as this simple example illustrates, the entanglement entropy and its fluctuations, characterized by the capacity of entanglement have no reason to be equal, except for a  specific coupling. This result is different from what happens for QFTs with a holographic gravity dual (without higher derivative terms) where the ratio $C_E/S_E$ turns out to be exactly equal to one \cite{DeBoer:2018kvc}.

\subsection{A lattice of oscillators and continuum limit} \label{sec:prelimQFT}
In this section we generalize our studies to a lattice of $N$ coupled harmonic oscillators, where the corresponding Hamiltonian is given by eq. \eqref{Nhamil}, in specific directions. Similar to the previous case, the ground state wave function is described by a factorized Gaussian in the corresponding normal mode space. To see this we recall that as shown in \cite{Srednicki:1993im} the matrix $K$ can be diagonalize by a real orthogonal similarity transformation, \textit{i.e.,} $K_D=UKU^T$.
Then writing the Hamiltonian in terms of the normal modes reduces the problem to a chain of decoupled quantum harmonic oscillators. Thus the total vacuum state can be written as
\begin{eqnarray}
\psi_0(\tilde{\textbf{x}})=\prod_{i=1}^N\left(\frac{\omega_i}{\pi}\right)^{1/4}\exp\left(-\frac{\omega_i\tilde{x}_i^2}{2}\right),
\end{eqnarray}
where $\tilde{x}_i$ and $\omega_i$ denote the normal coordinates and eigenfrequencies respectively. Further, in order to find the reduced density matrix we would like to express the above wave function in terms of the original variables $x_i$ in the position basis. It is relatively simple to show that in this case the above expression becomes
\begin{eqnarray}
\psi_0(\textbf{x})=\left(\frac{{\rm det}\;\Omega}{\pi^N}\right)^{1/4}\exp\left(-\frac{\textbf{x}^T.\Omega.\textbf{x}}{2}\right),
\end{eqnarray}
where $\Omega$ is the square root of $K$, \textit{i.e.,} 
\begin{eqnarray}\label{OmegaKD}
\Omega=U^TK_D^{1/2}U,\hspace*{2cm}(K_{D})_{ij}=\omega_i\;\delta_{ij}.
\end{eqnarray}
We would like to evaluate the reduced density matrix by tracing over the first $\widetilde{N}$ oscillators. To do so, we consider the following decomposition
\begin{eqnarray}\label{Omegadecom}
\Omega=\left(\begin{matrix}
A & B\\
B^T & C
\end{matrix}\right),
\end{eqnarray}
where $A$ is an $\widetilde{N}\times \widetilde{N}$ matrix and $C$ is an $(N-\widetilde{N})\times (N-\widetilde{N})$ matrix. Now it is straightforward to explicitly evaluate the reduced density matrix as\footnote{Note that $x$ now has $N-\widetilde{N}$ components, so we do not use the bold face notation which refers to eq. \eqref{xp}.} 
\begin{eqnarray}
\rho_{\rm red.}(x_{\widetilde{N}+1}, \cdots, x_N; x'_{\widetilde{N}+1}, \cdots, x'_N)=\mathcal{N}\exp\left(-\frac{x^T.\gamma.x+{x'}^T.\gamma.x'}{2}+x^T.\beta.x'\right),
\end{eqnarray}
where $\beta=\frac{B^TA^{-1}B}{2}$, $\gamma=C-\beta$ and $\mathcal{N}$ is a normalization factor. Moreover, as shown in \cite{Srednicki:1993im} the above expression can be rewritten as follows
\begin{eqnarray}\label{rhoredNosci}
\rho_{\rm red.}(z_{\widetilde{N}+1}, \cdots, z_N; z'_{\widetilde{N}+1}, \cdots, z'_N)=\mathcal{N}\;\prod_{i=\widetilde{N}+1}^N\exp\left(-\frac{z_i^2+{z'}_i^2}{2}+\tilde{\beta}_iz_iz'_i\right),
\end{eqnarray}
where 
\begin{eqnarray}
\tilde{\beta}=\gamma_D^{-1/2}V\beta V^T \gamma_D^{-1/2},\hspace*{1.5cm} z=W^T\gamma_D^{1/2}Vx.
\end{eqnarray}
Also $V$ and $W$ defined as the corresponding similarity transformations that diagonalize $\gamma$ and $\tilde{\beta}$ respectively, \textit{i.e.,} $\gamma_D=V\gamma V^T$ and $\tilde{\beta}_D=W\tilde{\beta} W^T$. Comparing eq. \eqref{rhoredNosci} to eq. \eqref{rho2} we see that the Renyi entropy can be calculated by summing the contribution for each of the remaining modes as follows
\begin{eqnarray}\label{Noscirenyi}
S_n=\frac{1}{1-n}\sum_{j=1}^{j_{\rm max}}\left(n\log(1-\xi_j)-\log(1-\xi_j^n)\right),
\end{eqnarray}
where $\xi_j=\frac{\tilde{\beta}_j}{1+(1-\tilde{\beta}_j^2)^{1/2}}$ and $j_{\max}={\min}(\widetilde{N}, N-\widetilde{N})$. Note that as long as we are
dealing with pure states, the spectrum of the reduced density matrices for $A$ and $\bar{A}$ is the same and hence $S_n(A)=S_n(\bar{A})$. Further, the corresponding expressions for entanglement entropy and capacity of entanglement are similar to eq. 
\eqref{2osciSECE} where again we should consider the contribution for each of the modes, \textit{i.e.},
\begin{eqnarray}\label{NosciSECE}
S_E&=&\sum_{j}\left(\frac{\xi_j}{\xi_j-1}\log\xi_j-\log(1-\xi_j)\right),\\
C_E&=&\sum_{j}\xi_j\left(\frac{\log\xi_j}{1-\xi_j}\right)^2.
\end{eqnarray}

We now wish to apply this approach to the problem of finding different entanglement measures in the continuum limit. For example considering a free scalar theory given by eq. \eqref{scalarhamil} and replacing the space continuum with a discrete mesh of lattice points the Hamiltonian can be transformed into a discrete counterpart as follows
\begin{eqnarray}\label{hamildis1}
H=\frac{1}{2}\sum_{i=1}^N \left(\pi_i^2+\left(\phi_{i+1}-\phi_{i}\right)^2+m^2\phi_{i}^2\right),
\end{eqnarray}
where without loss of generality we set the lattice spacing equal to unity, \textit{i.e.}, $\epsilon=1$. We see that the above expression and eq. \eqref{Nhamil} will be in
complete agreement if we choose
\begin{eqnarray}\label{hamildis2}
K_{ij}=(2+m^2)\delta_{i,j}-\left(\delta_{i+1,j}+\delta_{i,j+1}\right).
\end{eqnarray}
It is then possible to extract $\Omega$ using eq. \eqref{OmegaKD} and to use eqs. \eqref{Noscirenyi} and \eqref{NosciSECE} to determine the behavior of the entanglement measures. Generally, it is not possible to find the quantities analytically and thus in the following we will employ a numerical treatment.

\begin{figure}[h!]
	\begin{center}
\includegraphics[scale=0.87]{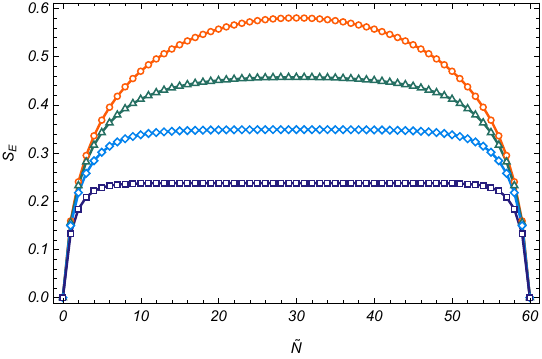}
  \hspace*{0.4cm}
\includegraphics[scale=0.87]{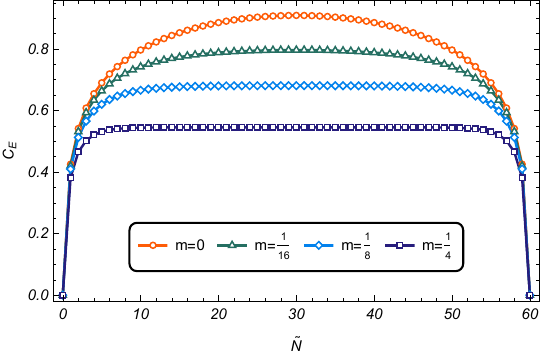}
  	\end{center}
	\caption{$S_E$ (left) and $C_E$ (right) as functions of $\widetilde{N}$ for several values of the mass parameter.}
	\label{fig:Nosci1}
\end{figure}

\begin{figure}[h!]
	\begin{center}
\includegraphics[scale=0.87]{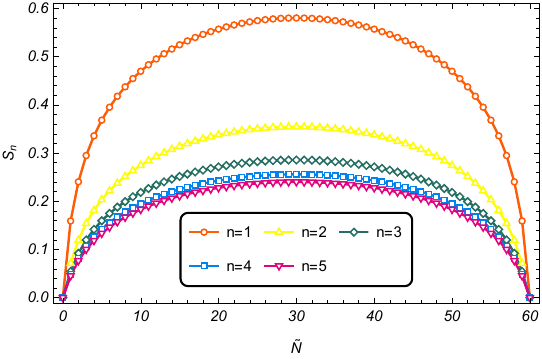}
  \hspace*{0.4cm}
\includegraphics[scale=0.87]{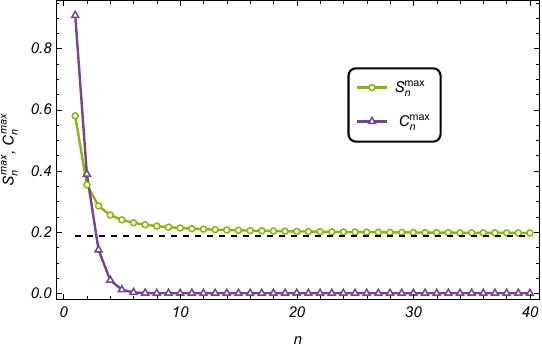}
  	\end{center}
	\caption{\textit{Left}: 	Renyi entropy as a function of $\widetilde{N}$ for several values of $n$. \textit{Right}: Maximum values of entanglement entropy and capacity of entanglement as a functions of the Renyi index. The dashed line indicates the asymptotic value of the Renyi entropy given by eq. \eqref{renyilargen}. In both plots we consider the massless regime.
	}
	\label{fig:Nosci2}
\end{figure}

The corresponding numerical results for different quantities are summarized in figures \ref{fig:Nosci1} and \ref{fig:Nosci2}. Note that we will mainly consider $N=60$, because the interesting qualitative features of the entanglement measures are independent of the total system size. Also this
choice facilitates a comparison to the analogous results for entanglement entropy in \cite{Srednicki:1993im}. In the left panel of figure \ref{fig:Nosci1}, the entanglement entropy is plotted as a function of $\widetilde{N}$ for several values of the mass parameter. Further, we show the behavior of capacity of entanglement for the same values of the parameters in the right panel of this figure. We can see that both measures have qualitatively similar behavior and decrease with $m$. Also in all cases $C_E$ is slightly larger than $S_E$. Notice that as long as we are dealing with pure states, the entanglement measures are symmetric, \textit{e.g.}, $S_{E}(A)=S_{E}(\bar{A})$ where $\bar{A}$ is the complement of $A$.  

In the left panel of figure \ref{fig:Nosci2}, we present the behavior of the Renyi entropy as a function of $\widetilde{N}$ for several values of $n$ in the massless regime. From this plot, one can infer that the qualitative features of the Renyi entropy are similar to $S_E$. Again, we see that the Renyi entropy is a decreasing function of $n$ such that the rate of change of $S_n$ is a monotonically decreasing function of the Renyi index and saturates from above to a constant in the large $n$ limit. Let us add that we found similar results for $C_n$, although we do not explicitly show the corresponding figures here. In order to investigate how these quantities approach the large $n$ limit, in the right panel of figure \ref{fig:Nosci2}, we plot the maximum values of $S_n$ and $C_n$ (which occurs at $\widetilde{N}=\frac{N}{2}$) as functions of the Renyi index. The dashed line indicates the asymptotic value of the Renyi entropy given by eq. \eqref{renyilargen} where the largest eigenvalue of the reduced density matrix can be evaluated numerically to be $\lambda_{\rm max}\sim 0.83$. Finally, the asymptotic behavior of $n$-th capacity is consistent with the large $n$ expansion, \textit{i.e.}, $C_{\infty}\sim \lim_{n\rightarrow \infty} n^2 \lambda_{\rm max}^{-n}\sim 0$.

\section{$C_E$ for harmonic systems in a squeezed state} \label{sec:harmonicsqueez}

In this section, we wish to return to our original problem which is computing the capacity of entanglement of a chain of coupled oscillators in squeezed states. Again, as a warm up problem, we first consider the case of a single pair of coupled harmonic oscillators and then generalize the problem to more involved case consisting of $N$ degrees of freedom in the next sections.

\subsection{Two coupled harmonic oscillators} \label{sec:2harmonicsqueez}

Let us begin with the case of two oscillators as in section \ref{sec:prelimtwoosci}, but now the overall system lying in a squeezed state given by eq. \eqref{squstate}. Again, we can write the total wave function as the product of the wave functions for the two individual oscillators as follows\footnote{Based on the results of \cite{Katsinis:2023hqn} in the following we will set $\phi_{s+}=\phi_{s-}=0$ without loss of generality.}
\begin{eqnarray}
\psi(x, t)&=&\left(\frac{{\Re}(w_+){\Re}(w_-)}{\pi^2}\right)^{\frac{1}{4}}\times\nonumber\\
&&\hspace*{-0.14cm}\exp\hspace*{-0.09cm}\left(\hspace*{-0.09cm}-\frac{w_+(x_+-x_{0+})^2+w_-(x_--x_{0-})^2}{2}+\hspace*{-0.09cm}ip_{0+}(x_+-x_{0+})+\hspace*{-0.09cm}ip_{0-}(x_--x_{0-})\right)\hspace*{-0.09cm},
\end{eqnarray}
where $x_{\pm}$ were defined in eq. \eqref{xpm} and we have also introduced the new variables $x_{0\pm}=\frac{x_{01}\pm x_{02}}{\sqrt{2}}$ and $p_{0\pm}=\frac{p_{01}\pm p_{02}}{\sqrt{2}}$. 
The corresponding reduced density matrix can be found by rewriting the wave function in terms of the original coordinates and integrating out the $x_1$ variable. The calculation follows straightforwardly from the considerations of the previous section. We have then
\begin{eqnarray}
\rho_2(x_2; x'_2)=\left(\frac{{\rm Re}(\gamma)-\beta}{\pi}\right)^{1/2}\exp\left(-\frac{\gamma y_2^2+\gamma^* {y_2'}^2}{2}+\beta {y}_2y'_2+ip_{02}(y_2-y_2')\right),
\end{eqnarray}
where we have defined $y_2=x_2-x_{02}$, $y_2'=x_2'-x'_{02}$ and
\begin{eqnarray}
\beta=\frac{|w_+-w_-|^2}{4{\rm Re}\left(w_++w_-\right)},\hspace*{2cm}\gamma=\frac{|w_++w_-|^2+4w_+w_-}{4{\rm Re}\left(w_++w_-\right)}.
\end{eqnarray}
Interestingly, although $\gamma$ is complex, the eigenvalues of the reduced density matrix depend only on the real part of this parameter \cite{Katsinis:2023hqn}. Indeed, as shown in this reference the calculation proceeds essentially as the ground state case and the resulting spectrum is given by
\begin{eqnarray}\label{pkxisqu}
p_k=(1-\xi)\xi^k,
\end{eqnarray}
where $\xi=\frac{\beta}{{\rm Re}(\gamma)+\alpha}$ and $\alpha=\sqrt{{\rm Re}(\gamma)^2-\beta^2}$.
Again we can evaluate different entanglement measures using eqs. \eqref{2oscirenyi} and \eqref{2osciSECE}. It is worthwhile to mention that based on eq. \eqref{squstate1} the spectrum of the reduced density matrix depends on time and thus we expect that the measures are also time-dependent. 

\begin{figure}[h!]
	\begin{center}
\includegraphics[scale=0.87]{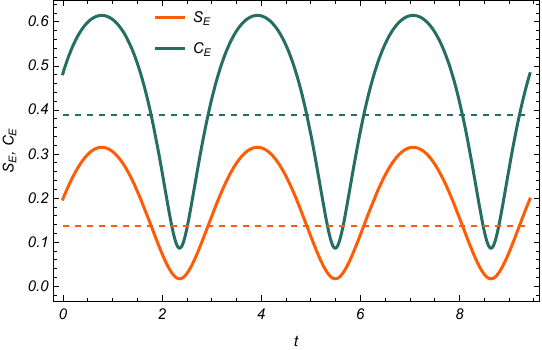}
  \hspace*{0.4cm}
\includegraphics[scale=0.87]{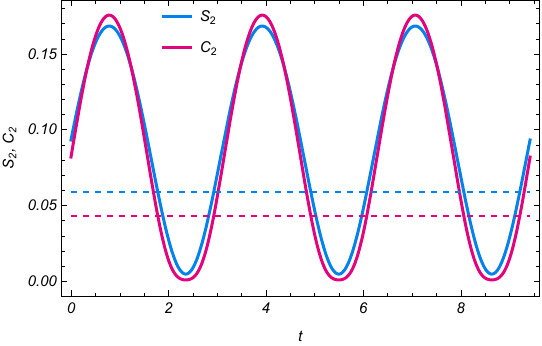}
  	\end{center}
	\caption{\textit{Left}:	Evolution of entanglement entropy and capacity of entanglement. \textit{Right}: $S_n$ and $C_n$ as a function of time for $n=2$. In both panels we set $\omega_-=2\omega_+=2$, $z_+=0.5$ and $z_-=0$. Also the dashed lines are the corresponding results for the ground state where both squeezing parameter vanish.
	}
	\label{fig:twoosci1squ}
\end{figure}

\begin{figure}[h!]
	\begin{center}
\includegraphics[scale=0.99]{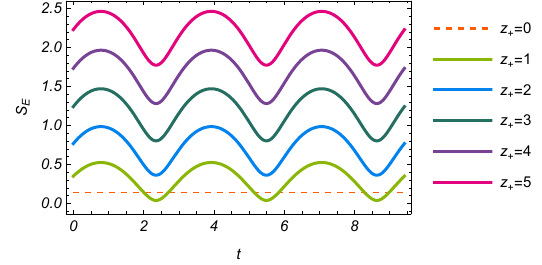}
\includegraphics[scale=0.77]{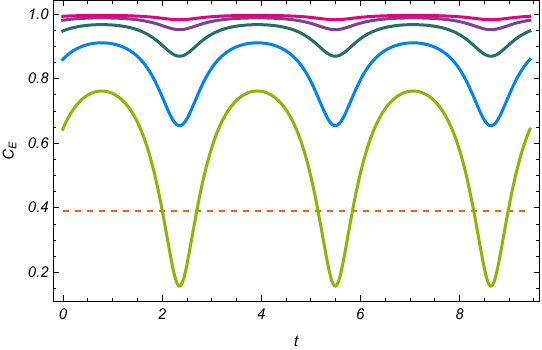}
  	\end{center}
	\caption{Evolution of entanglement entropy (left) and capacity of entanglement (right) for several values of $z_+$ for $\omega_-=2\omega_+=2$ and $z_-=0$.	}
	\label{fig:twoosci2squ}
\end{figure}

With these tools in hand, let us examine the dependence of the measures on the squeezed parameter in more detail. The numerical results for different quantities are summarized in figures \ref{fig:twoosci1squ} and \ref{fig:twoosci2squ}.
Figure \ref{fig:twoosci1squ} shows the time evolution of entanglement measures when only the symmetric mode is squeezed. From these plots, one can infer that different measures exhibit qualitatively similar behaviors. Clearly, in this case the evolution is periodic with period $\tau_+\equiv\frac{T_+}{2}=\frac{\pi}{\omega_+}$. We see that $C_E$ reaches it maximum values at times $t_{\rm max}=(4k+1)\frac{\tau_+}{4}$ where $k$ is an integer.
Interestingly, the capacity of entanglement is always greater than the entanglement entropy such that the difference of these quantities becomes maximal when they reach their maximum values. Moreover, the right panel shows that for larger values of $n$ this difference becomes less pronounced. 

Now we proceed further by examining in more detail the $z$ dependence of the entanglement measures, as shown in figure \ref{fig:twoosci2squ}. The key observation to
note here is that while both $S_E$ and $C_E$ are monotonically increasing functions of the squeezing parameter, only the entanglement entropy goes on to grow indefinitely as we increase $z$. Indeed, the right panel illustrates that $C_E$ saturates to unity in the large $z$ limit. To gain some insights into this behavior, let us turn our attention to the computation of the mean quantities which is defined as $\overline{\mathcal{M}}\equiv\frac{1}{T}\int_0^T\mathcal{M}(t)dt$. Before examining the full $z$ dependence of the mean capacity of entanglement, we would like to study its asymptotic behaviors in small and large $z$ limit. In \cite{Katsinis:2023hqn} it was shown that in these limits the expansions for the mean entanglement entropy become 
\begin{eqnarray}\label{expandse}
\overline{S_E}=\Bigg\{ \begin{array}{lcr}
S_{E0}-\frac{1}{16}\left(1+\frac{1+4\xi_0+\xi_0^2}{1-\xi_0^2}\ln \xi_0\right)z_++\cdots, &\,\,\,z_+\ll 1,\\
\frac{z_+}{2}+1-3\ln2+\ln\left(\sqrt{\frac{\omega_+}{\omega_-}}+\sqrt{\frac{\omega_-}{\omega_+}}\right)+\mathcal{O}(e^{-z_+}), &\,\,\,z_+\gg 1,
\end{array}
\end{eqnarray}
where $S_{E0}$ and $\xi_0$ denote the corresponding vacuum values with $z_+=0$. This shows that for a large squeezing parameter the mean entanglement entropy grows linearly with $z$. A similar derivation holds in the present case and it is also easy to find that the expression for the mean capacity of entanglement reduces to 
\begin{eqnarray}\label{expandcapa}
\overline{C_E}=\Bigg\{ \begin{array}{lcr}
C_{E0}+\frac{1}{8}\left(1+2\frac{1+3\xi_0+\xi_0^2}{1-\xi_0^2}\ln \xi_0+\frac{1}{2}\frac{1+6\xi_0+\xi_0^2}{(1-\xi_0^2)^2}\left(\ln \xi_0\right)^2\right)z_+^2+\cdots, &\,\,\,z_+\ll 1,\\
1-\mathcal{O}(e^{-2z_+}), &\,\,\,z_+\gg 1,
\end{array}
\end{eqnarray}
which shows that in the large $z$ limit, the capacity of entanglement saturates to unity. 

\begin{figure}[h!]
	\begin{center}
\includegraphics[scale=0.87]{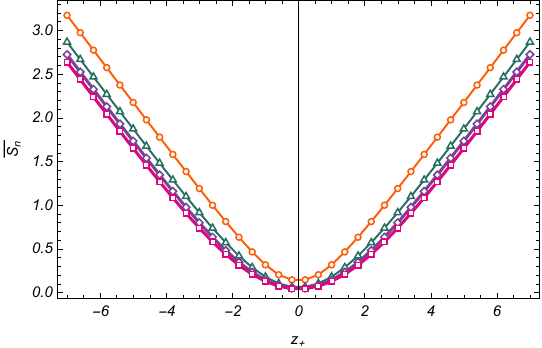}
\hspace*{0.4cm}
\includegraphics[scale=0.87]{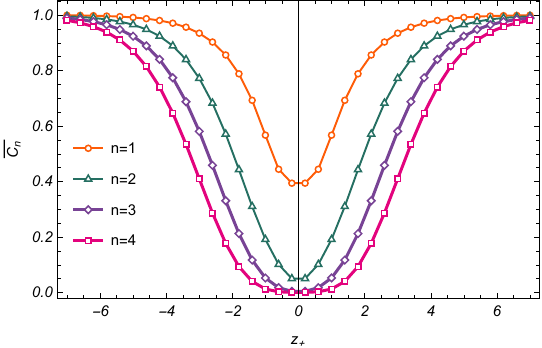}
  	\end{center}
	\caption{Mean values of the Renyi entropy (left) and $n$-th capacity of entanglement (right) as functions of the squeezing parameter for several values of the Renyi index. The asymptotic behavior of $\overline{C_n}$ in large $z$ limit is almost independent of $n$. Here we set $\omega_-=2\omega_+=2$ and $z_-=0$.
	}
	\label{fig:twoosci4squ}
\end{figure}

The numerical results for the Renyi entropy and $n$-th capacity of entanglement as functions of the squeezing parameter for different values of the Renyi index with specific values of $\omega_\pm$ are summarized in figure \ref{fig:twoosci4squ}. Note that the case of $n=1$ corresponds to $\overline{S_E}$ and $\overline{C_E}$. In this case our numerical results coincide with the analytical expansions given by eqs. \eqref{expandse} and \eqref{expandcapa}. Interestingly, we see that in the large $z$ limit $\overline{C_E}/\overline{S_E} \ll 1$ which shows that the reduced density matrix becomes more and more maximally mixed as one increases the squeezing parameter.
The left panel illustrates that all curves for $\overline{S_n}$ have the same behavior in large $n$ limit. In particular, one gets asymptotic linear growth with the same slope for different values of $n$ and $z$. Moreover, the right panel also shows that the asymptotic behavior of the mean value of the $n$-th capacity of entanglement in large squeezing limit is almost independent of the Renyi index. Indeed, it is straightforward to show that the corresponding scaling of $\overline{S_n}$ and $\overline{C_n}$ with $z$ at leading order is the same as eqs. \eqref{expandse} and \eqref{expandcapa} respectively.  

To close this subsection, let us comment on extending this discussion to cases where both modes are squeezed. Indeed, in this case the evolution of the entanglement measures is in general not periodic, as the ratio of the frequencies of the two modes may be irrational. Again we found similar results for small and large squeezing parameter, although we do not explicitly show the corresponding figures here.

\subsection{A chain of oscillators} \label{sec:Nharmonicsqueez}

In this section we again consider the system of $N$ coupled harmonic oscillators which was defined in eq. \eqref{Nhamil} to compute several entanglement measures, but now for the squeezed states given by eq. \eqref{squstate}. The main question of interest is what are the additional technicalities involved in computing different measures in squeezed states compared to the vacuum state. In particular, we would like to investigate to what extent the squeezing parameter modifies the behavior of capacity of entanglement. Here, it is worth mentioning that, the influence of this parameter on the entanglement entropy was studied in \cite{Katsinis:2023hqn} and we will follow the discussion there closely. Indeed,  the spectrum of the reduced density matrix corresponding to a squeezed state can be evaluated as detailed in this reference. These authors proposed three equivalent scenarios to compute the eigenvalues of $\rho_A$. In what follows we briefly review one of the approaches used in the reference which is fairly simple. To do so, assume that all the corresponding normal modes lie in a squeezed state whose wave function is given as follows
\begin{eqnarray}
\psi(\textbf{x})=\left(\frac{{\rm det}\;{\rm Re}\;(\Omega)}{\pi^N}\right)^{1/4}\exp\left(-\frac{(\textbf{x}-\textbf{x}_0).\Omega.(\textbf{x}-\textbf{x}_0)}{2}+i\textbf{p}_0.(\textbf{x}-\textbf{x}_0)-i\sum_{j}\phi_{sj}\right),
\end{eqnarray}
where $\Omega$ was defined in eq. \eqref{OmegaKD}, but now $\omega_i$'s are given by the application of eq. \eqref{squstate1} for each normal mode and hence it is a complex
symmetric matrix. It is straightforward to show that in this case the reduced density matrix for the remaining $(N-\widetilde{N})$ oscillators becomes
\begin{eqnarray}\label{rhoredNoscisqu1}
\rho_{\rm red.}(x_{\widetilde{N}+1}, \cdots; x'_{\widetilde{N}+1}, \cdots)\hspace*{-0.1cm}=\hspace*{-0.1cm}\left(\frac{{\rm det}\,{\rm Re}(\gamma-\beta)}{\pi^{N-\widetilde{N}}}\right)^{\hspace*{-0.1cm}\frac{1}{2}}\hspace*{-0.1cm}\exp\hspace*{-0.1cm}\left(\frac{-y.\gamma.y-y'.\gamma^*.y'}{2}+y'.\beta y+ip_{0}.(y_2-y'_{2})\right),
\end{eqnarray}
where
\begin{eqnarray}\label{rhoredNoscisqu2}
\beta=\frac{1}{2}B^{\dagger}{\rm Re}(A)^{-1}B,\hspace*{1.5cm}\gamma=C-\frac{1}{2}B^{T}{\rm Re}(A)^{-1}B,
\end{eqnarray}
and $y=x-x_0$. Note that we consider the same decomposition for $\Omega$ as in eq. \eqref{Omegadecom}. Also note that here $\gamma$ is a real symmetric matrix and $\beta$ is a Hermitian matrix. Indeed, this result is different from what happens for the ground state where $\beta$ is a real and symmetric matrix. Hence in this case it cannot be diagonalized via a real orthogonal transformation. Fortunately, as shown in \cite{Katsinis:2023hqn} for our purposes a general solution is not required and in order to find the eigenvalues of eq. \eqref{rhoredNoscisqu1}, it is sufficient to compute the spectrum of a simpler matrix $\tilde{\Omega}$ defined by
\begin{eqnarray}\label{tildeOmegad}
\tilde{\Omega}={\rm Re}(\Omega)^{-1}\left(\begin{matrix}
-{\rm Re}(A) & i{\rm Im}(B)\\
-i{\rm Im}(B)^T & {\rm Re}(C)
\end{matrix}\right).
\end{eqnarray}
The relation between the two spectra is $\xi_j=\frac{\tilde{\xi}_j-1}{\tilde{\xi}_j+1}$ where the normalization condition for the density matrix allows us to neglect $\tilde{\xi}_j<1$. Further, the eigenstates can be written in terms of the Hermite polynomials. We skip over the details of the calculation and we refer the interested reader to \cite{Katsinis:2023hqn} for further details. Having the corresponding eigenvalues, the entanglement measures can be found using the same formulas as in the previous section, \textit{e.g.}, eq. \eqref{NosciSECE}. Now we are equipped with all we need to study the behavior of the entanglement spectrum and thereby other related quantities in a squeezed state. To do so, we employ a numerical treatment in the next section. 

Before we proceed further, we would like to study the asymptotic behaviors of the entanglement measures in large squeezing limit. Indeed, this study plays an important role in our analysis in the next section. Here for simplicity we restrict our analysis to a specific case where all modes lie in a squeezed state with the same squeezing
parameter which is very large, \textit{i.e.}, $z\gg 1$. Of course, a similar analysis has been previously done for the entanglement entropy in \cite{Katsinis:2023hqn}. Indeed, as shown in this reference in the large $z$ limit the corresponding eigenvalues can be written as
\begin{eqnarray}\label{largezeigen}
\xi_j=1-e^{-z}\xi^{(1)}_j+\cdots,
\end{eqnarray}
where $\xi^{(1)}_j$ are some $z$ independent positive coefficients. Using the above expression one finds the following expansion for the entanglement entropy
\begin{eqnarray}\label{largezSE}
S_E={\rm min}(\widetilde{N}, N-\widetilde{N})(z+1)-\sum_{j=1}^{j_{\rm max}}\log \xi^{(1)}_j-\mathcal{O}\left(e^{-z}\right).
\end{eqnarray}
Thus in the large $z$ limit the entanglement entropy has a linear dependence on the squeezing parameter. In a similar manner, one can show that
\begin{eqnarray}\label{largezRE}
S_n={\rm min}(\widetilde{N}, N-\widetilde{N})\left(z+\frac{\log n}{n-1}\right)-\sum_{j=1}^{j_{\rm max}}\log \xi^{(1)}_j+\mathcal{O}\left(e^{-z}\right),
\end{eqnarray}
which shows that the Renyi entropy also has a linear behavior in this limit. Further, the corresponding expression for the capacity of entanglement can be determined in a similar way and the result is 
\begin{eqnarray}\label{largezCE}
C_E={\rm min}(\widetilde{N}, N-\widetilde{N})-\mathcal{O}\left(e^{-2z}\right).
\end{eqnarray}
Thus asymptotically, it approaches a constant value which is exactly the number of degrees of freedom of the smaller subsystem. Indeed, based on the above results we see that at leading order all the measures are time-independent and proportional to the volume of the smaller subsystem. Again, we see that in this limit $C_E\ll S_E$ which shows that the reduced density matrix becomes more and more maximally mixed as one increases the squeezing
parameter. It is also easy to show that in this case the expansion for $C_n$ is the same as $C_E$.
In the next section, we numerically evaluate the entanglement measures for a free scalar theory in a squeezed state which enables
us to more explore the validity of these expressions.  

\section{$C_E$ for a massless scalar in squeezed states} \label{sec:QFTsqueez}

In this section we proceed our previous analysis in a scalar theory, whose Hamiltonian is given by eq. \eqref{hamildis1}, to more investigate the behavior of the capacity of entanglement in a squeezed state. First, we provide a simple analysis where just a single mode is squeezed and examine the various regimes in the evolution of entanglement measures. Next, we will extend this study to a more general case where all the normal modes lie in a squeezed state which enables us to directly extract some interesting features of the $z$ dependence of the capacity of entanglement.

\subsection{Squeezing a single mode} \label{sec:2harmonicsqueezsingle}
\begin{figure}[h!]
	\begin{center}
\includegraphics[scale=0.59]{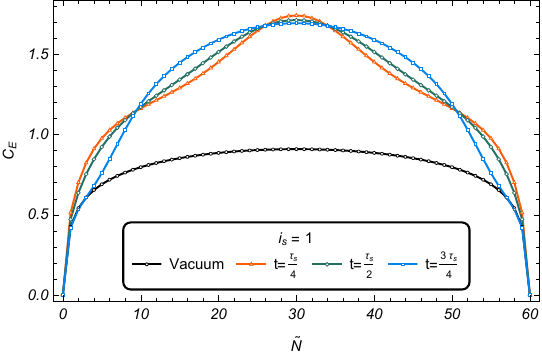}
  \includegraphics[scale=0.59]{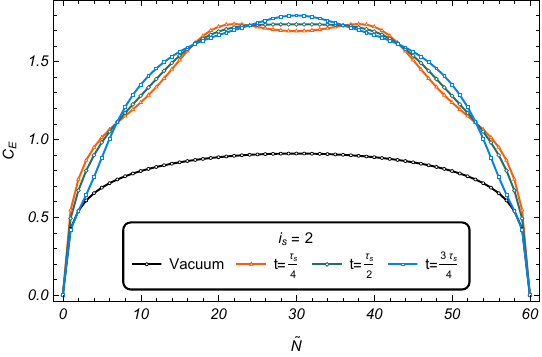}
\includegraphics[scale=0.59]{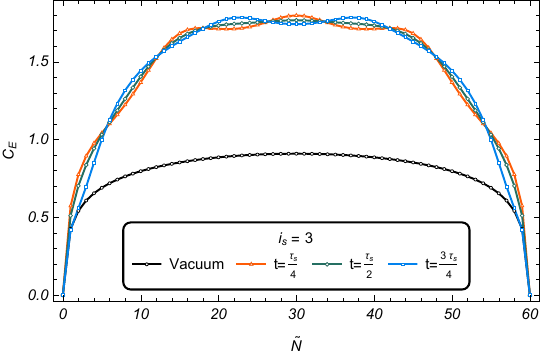}
  	\end{center}
	\caption{The capacity of entanglement as a function of $\widetilde{N}$ for several time slices when only a single mode
has been squeezed with $z=3$. The black markers correspond to the vacuum state. 	}
	\label{fig:Ctmodes}
\end{figure}

To begin, we consider the system lying in a state where only one normal mode is squeezed and thus its wave function is given by eq. \eqref{squstate}. Also the remaining modes are put in their ground states. In the following we will use $i_s$ to denote the index of the squeezed mode. Further, based on the results of section \ref{sec:2harmonicsqueez} we expect that the evolution of the measures is periodic with period $\tau_s\equiv\frac{T_s}{2}=\frac{\pi}{\omega_s}$ where $\omega_s$ is the normal frequency of the corresponding squeezed mode. The examples depicted in figure \ref{fig:Ctmodes} exhibit this behavior for the capacity of entanglement. In the figure, $C_E$ as a function of $\widetilde{N}$ is presented for several values of $i_s$ with $z=3$. The black markers correspond to the vacuum state.
Based on this figure, it is evident that squeezing mostly increases $C_E$ in comparison to that in the
vacuum state. Nevertheless, we notice that for $i_s=1$ and $t=0.75 \,\tau_s$ the capacity of entanglement becomes smaller than that of the vacuum state. It is expected that this behavior also happens for other values of $z$ and $i_s$. Further, the oscillations of $C_E$ can be traced back to the trigonometric functions appearing in the wave function corresponding to the squeezed state. The larger $i_s$, the shorter the period of oscillation. Also as this parameter increases, the amplitude of oscillation becomes smaller. Moreover, one can see that in this regime the oscillation seems more or less the same independent of $i_s$. Let us add that similar behavior was found for the entanglement entropy in \cite{Katsinis:2023hqn}. 
\begin{figure}[h!]
	\begin{center}
\includegraphics[scale=0.87]{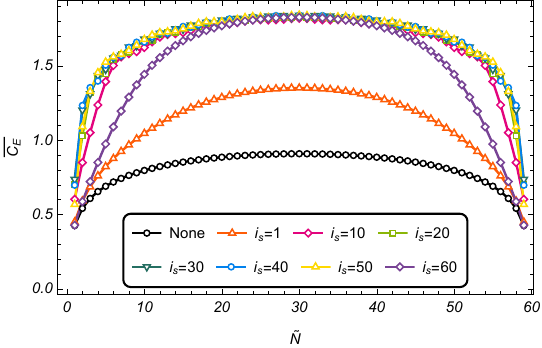}
  \hspace*{0.4cm}
\includegraphics[scale=0.87]{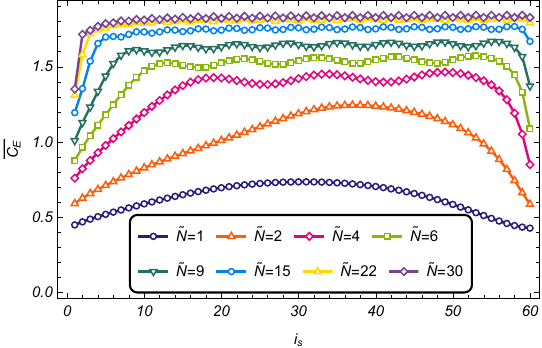}
  	\end{center}
	\caption{Mean capacity of entanglement as a function of $\widetilde{N}$ (left) and $i_s$ (right) for several values of the parameters when a single mode has been
squeezed with $z=3$.
	}
	\label{fig:SCbarnmodes}
\end{figure}

To gain some insights into this behavior, let us turn our attention to the computation of the mean entanglement measures in this setup. The corresponding results for the mean capacity of entanglement are summarized in figure \ref{fig:SCbarnmodes}. The left panel illustrates $\overline{C_E}$ as a function of $\widetilde{N}$ for several values of the index of the squeezed mode. We see that the mean capacity of entanglement is always larger than that in the vacuum state of the system. Further, the right panel shows this quantity as a function of $i_s$ for several divisions of the system in two subsystems. We note a number of key features: First, $\overline{C_E}$ is not monotonous with the index of the squeezed mode. Indeed, for small values of $\widetilde{N}$ the mean capacity of entanglement first increases approximately linearly with $i_s$ and then after some fluctuations decreases to a constant value. Note that the intermediate fluctuations become less pronounced in the large $\widetilde{N}$ limit. Second, for large values of $\widetilde{N}$ the mean capacity of entanglement first increases very sharply with $i_s$ and then suddenly saturates to a constant value. 
%

\begin{figure}[h!]
	\begin{center}
\includegraphics[scale=0.87]{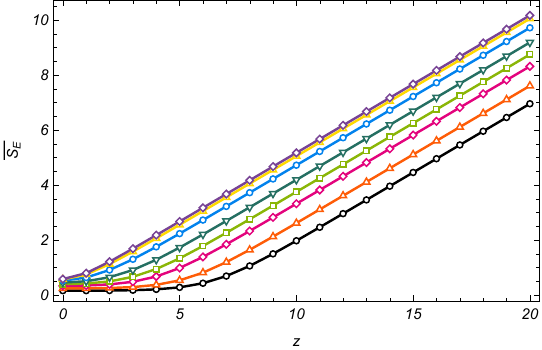}
  \hspace*{0.2cm}
\includegraphics[scale=0.87]{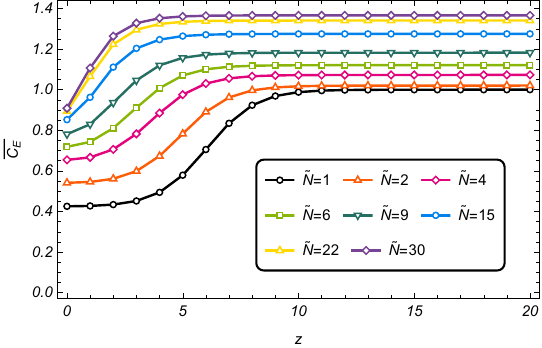}
%
  	\end{center}
	\caption{Entanglement entropy (left) and capacity of entanglement (right) as functions of $z$ for $i_s=1$ and several values of $\widetilde{N}$.
	}
	\label{fig:SCbarzmode130}
\end{figure}

\begin{figure}[h!]
	\begin{center}
\includegraphics[scale=0.87]{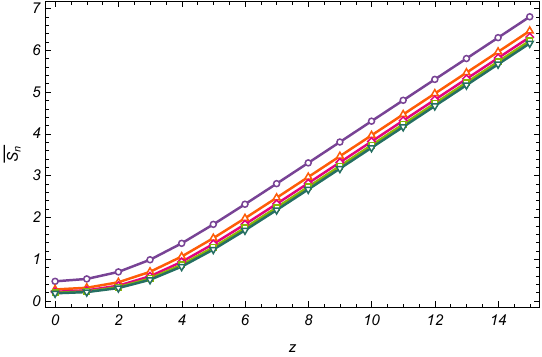}
  \hspace*{0.2cm}
\includegraphics[scale=0.87]{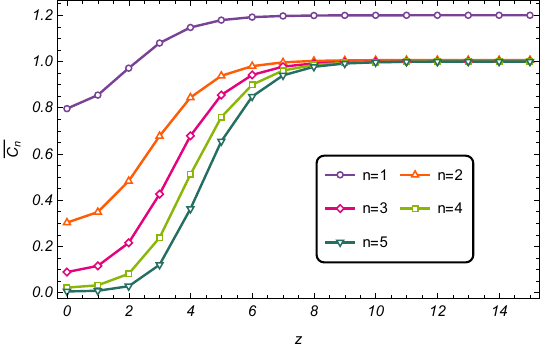}
  	\end{center}
	\caption{Mean values of the Renyi entropy (left) and $n$-th capacity of entanglement (right) as functions of the squeezing parameter for several values of the Renyi index with $i_s=1$ and $\widetilde{N}=10$.
	}
	\label{fig:nthSCbarzmode1}
\end{figure}

To close this subsection, we examine the $z$ dependence of mean entanglement measures in figures \ref{fig:SCbarzmode130} and \ref{fig:nthSCbarzmode1}. We focus our analysis
on the special case of $i_s=1$ because the interesting qualitative features of the measures do not depend
strongly on which mode is squeezed. Note that we have also included the corresponding results for the mean entanglement entropy, which was previously reported in \cite{Katsinis:2023hqn}, to allow for a meaningful comparison between the different measures. Based on figure \ref{fig:SCbarzmode130}, we observe that both $\overline{S_E}$ and $\overline{C_E}$ are monotonically increasing functions of the squeezing parameter, as expected. Further, the left panel illustrates that for large squeezing, the entanglement entropy has a linear dependence on $z$. Remarkably, from the right panel we can deduce that the mean capacity of entanglement saturates to a constant value in this regime. Hence, we expect that the corresponding reduced density matrix becomes more and more maximally mixed as one increases the squeezing parameter. Moreover, the saturation value is a monotonically increasing function of both $\widetilde{N}$ and $i_s$. 

Figure \ref{fig:nthSCbarzmode1} illustrates the behavior of mean values of the Renyi entropy and $n$-th capacity of entanglement as functions of the squeezing parameter for several values of the Renyi index. The left panel shows that $\overline{S_n}$ exhibits a linear dependence on the squeezing parameter with the same slope for different values of $n$. Further, in the right panel we see that $\overline{C_{n}}$ with $n\geq 2$ saturate from below to unity independent of the Renyi index. Indeed, for larger values of $n$, the mean value of the $n$-th capacity of entanglement vanishes initially but then rapidly rises to the final constant value corresponding to the large squeezing regime. 
%
%

\subsection{Squeezing all modes} \label{sec:2harmonicsqueezallmode}

We turn now to a more realistic case where all modes lie in a squeezed state with the same squeezing parameter. As in the previous section, unlike the case where we had squeezed a single mode, the evolution of entanglement measures are not necessarily periodic. Moreover, as we have shown in section \ref{sec:Nharmonicsqueez}, the large $z$ behavior of the measures are completely independent of time. In particular, in this regime the scaling of the capacity of entanglement is not sensitive to the details of the time evolution of the reduced density matrix. 

\begin{figure}[h!]
	\begin{center}
\includegraphics[scale=0.59]{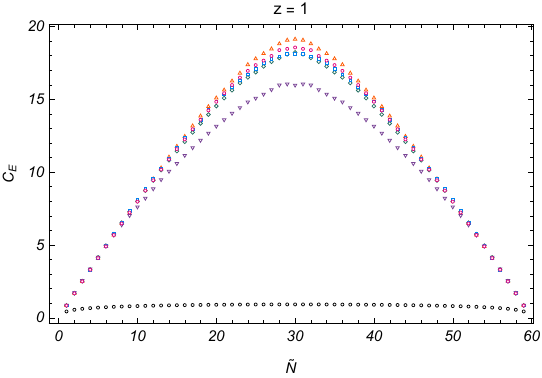}
\includegraphics[scale=0.59]{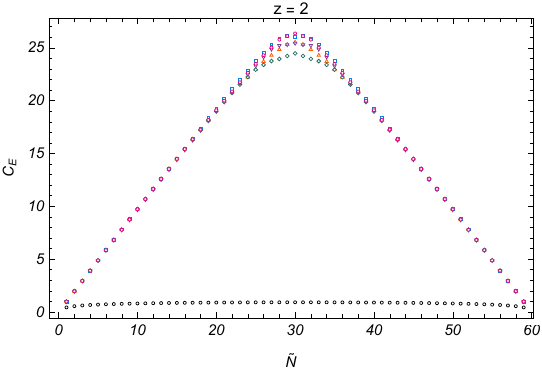}
\includegraphics[scale=0.59]{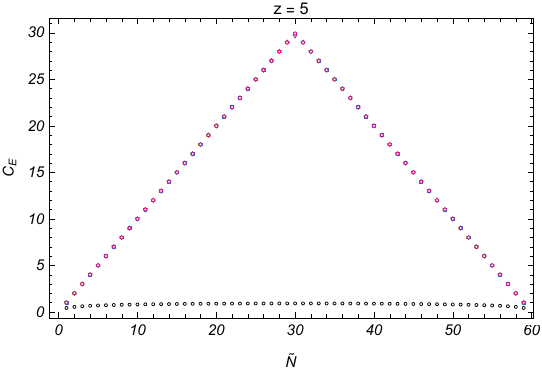}
  	\end{center}
	\caption{Capacity of entanglement for various random instants as functions of $\widetilde{N}$ when all modes lie in a squeezed state with the same squeezing parameter. The black markers correspond to the vacuum state. 
	}
	\label{fig:Crandom}
\end{figure}

\begin{figure}[h!]
	\begin{center}
\includegraphics[scale=0.87]{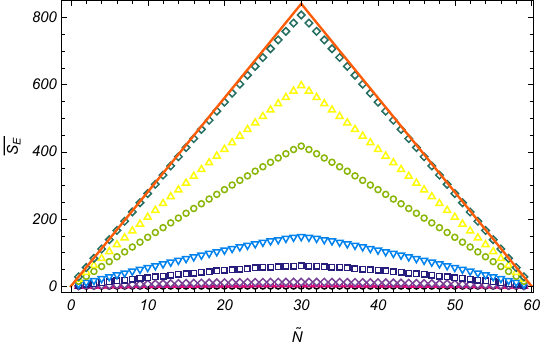}
  \hspace*{0.2cm}
\includegraphics[scale=0.87]{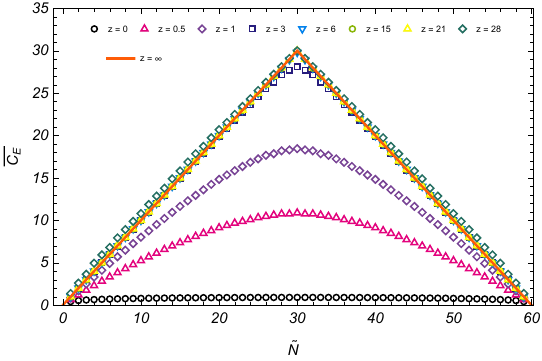}
  	\end{center}
	\caption{Mean entanglement entropy (left) and mean capacity of entanglement (right) as functions of $\widetilde{N}$ for several values of $z$. The red lines indicate the asymptotic behavior in large $z$ limit given by eqs. \eqref{largezSE} and \eqref{largezCE}. The mean has been calculated as the average of 300 random times.
	}
	\label{fig:SCbarallmode}
\end{figure}

The corresponding numerical results are summarized in figures \ref{fig:Crandom} and \ref{fig:SCbarallmode}. The capacity of entanglement for various random instants as function of $\widetilde{N}$ when all modes lie in a squeezed state with the same squeezing parameter is depicted in figure \ref{fig:Crandom}. Once again, the black markers correspond to the vacuum state. Note that the qualitative behavior of the capacity of entanglement is more or less the same independent of the random instant that we have chosen. Clearly, $C_E$ is a monotonically increasing function of the squeezing parameter. Further, as we increase the squeezing parameter the scaling of the capacity of entanglement tends asymptotically to eq. \eqref{largezCE} as expected. Let us recall that according to this equation the leading term is time-independent and thus the fluctuations of the capacity of entanglement with time decrease as the squeezing parameter increases.

To get a better understanding of the results, we compare the behavior of mean entanglement entropy and mean capacity of entanglement in figure \ref{fig:SCbarallmode}. The markers in this figure show the numerical results which in the large $z$ limit coincide with the leading term of the asymptotic expansions given by eqs. \eqref{largezSE} and \eqref{largezCE} (represented by the continuous red line). Further, based on our numerical results we see that at leading order $\overline{S_E}$ and $\overline{C_E}$ are proportional to the volume of the smaller subsystem. Indeed, for an infinite chain of oscillators in the continuum limit we have
\begin{eqnarray}\label{largezCE1}
\overline{S_E}=\frac{z\ell}{\epsilon}+\cdots,\hspace{2cm}\overline{C_E}=\frac{\ell}{\epsilon}+\cdots,
\end{eqnarray}
where $\ell\equiv \widetilde{N}\epsilon$ and $L\equiv N\epsilon\rightarrow \infty$. Thus in this limit $\frac{\overline{S_E}}{\overline{C_E}}=z\gg 1$ which is a feature of an ``almost'' maximally mixed state. Moreover, the measures are both sensitive to the UV cutoff, but the ratio is finite. We conclude that at least for the free massless scalar field theory, the ratio of the leading terms is scheme independent. Of course, the numerical results depicted in the left panel for the mean entanglement entropy are consistent with the Page's argument that in an arbitrary quantum state the entropy is close to maximal. Hence we expect entanglement entropy to scale with the volume (instead of the area) of the entangling region. 

The right panel shows that this special behavior also holds for capacity of entanglement. In particular, we see that $\overline{C_E}$ is always continuous and varies smoothly. This feature contrasts with the holographic results for the capacity of entanglement corresponding to the black hole evaporation process as previously noted in \cite{Kawabata:2021hac,Kawabata:2021vyo}. Of course, their results should be hold in a strongly coupled theory with a large number of degrees of freedom which is completely different from our free scalar model. Thus there is no a priori reason to expect that the results should agree in these cases. Interestingly enough, based on eq. \eqref{largezCE1} we see that in large squeezing limit the relationship $C_E=S_E$ is completely broken. Indeed, as proposed in \cite{DeBoer:2018kvc} perhaps such a relation in QFTs is a hint of a dual gravitational interpretation. Therefore, at least in this regime the corresponding squeezed states cannot have a solution of a classical gravity theory as a holographic dual. This is the main clue that forbids us to compare our results with some previous studies on capacity of entanglement in the context of holography. Let us emphasize that we found similar results for $\overline{S_n}$ and $\overline{C_n}$ which are consistent with the asymptotic expansions we reported in the previous section, although we do not explicitly show the corresponding figures here.


\section{Conclusions and discussions}\label{sec:diss}

In this paper, we explored the evolution and scaling of the entanglement measures in squeezed states which are the most general Gaussian states. We have mainly studied the behavior of capacity of entanglement, the quantum information theoretic counterpart of heat capacity, for a specific harmonic model which is a discrete counterpart of a $(1+1)$-dimensional free scalar field theory. To gain some intuition for the problem, we began by studying the simple case of a pair of harmonic oscillators. In this case we have studied different aspects of the capacity of entanglement in various setups and with different parameters numerically.

In particular, the time evolution of entanglement measures when only a single mode has been squeezed is periodic with period equal to half the period of the corresponding mode. Further, the capacity of entanglement is always greater than the entanglement entropy such that the difference of these quantities becomes maximal when they reach their maximum values. A key observation to note here is that while both $S_E$ and $C_E$ are monotonically increasing functions of the squeezing parameter, only the entanglement entropy
goes on to grow indefinitely as we increase $z$. Indeed, the capacity of entanglement saturates to unity in the large $z$ limit. Hence, we see that in this limit $\frac{C_E}{S_E}\ll 1$ which shows that the corresponding reduced density matrix becomes more and more maximally mixed as one increases the squeezing parameter. Moreover, the asymptotic behavior of
the mean value of the Renyi entropy and $n$-th capacity of entanglement in large squeezing limit is almost independent
of the Renyi index. In addition, considering more general cases where both modes are squeezed, although the evolution of the entanglement measures is in general not
periodic, the qualitative features of the results are the same. 

We have also extended these studies to the system of $N$ coupled harmonic oscillators to investigate to what extent the squeezing parameter modifies the behavior of the entanglement measures including the capacity of entanglement. In particular, we have shown that where all modes lie in a squeezed state the measures are time-independent and proportional to the volume of the smaller subsystem in large squeezing limit. Again, we have found that in this limit the ratio of capacity over entropy is negligible and thus the reduced density matrix behaves as an ``almost'' maximally mixed state.

In order to gain further insights into certain properties of these quantities in squeezed states, we have also considered a scalar field theory, which is the continuum counterpart of our harmonic chain. In this case when only a single mode has been squeezed, we found a number of key features: First, the mean capacity of entanglement is not monotonous with the index of the squeezed mode such that for large values of $\widetilde{N}$, it first increases very sharply with $i_s$ and then suddenly saturates to a constant value. Second, we observed that $C_E$ is a monotonically increasing function of the squeezing parameter and saturates to a constant value
in the large $z$ limit. Further, the mean Renyi entropy exhibits a linear dependence on the squeezing parameter with the same
slope for different values of the Renyi index. Also we have shown that the mean value of $n$-th capacity saturates from below to unity independent of $n$. Indeed, for larger values of the Renyi index, $\overline{C_n}$ vanishes initially but then rapidly rises to a constant value corresponding to the large squeezing regime.

Once again, considering the more realistic case where all modes lie in a squeezed state, the large $z$ behavior of the measures are completely independent of time. In particular, in this regime the scaling of the capacity of entanglement is not sensitive to the details of the time evolution of the reduced density matrix. We have also found numerically that at leading order $C_E$ is proportional to the volume of the smaller subsystem which is consistent with asymptotic scaling given by eq. \eqref{largezCE}. Interestingly, for an infinite chain of oscillators in the continuum limit the corresponding behaviors of $S_E$ and $C_E$ are given by eq. \eqref{largezCE1} which shows that the reduced density matrix becomes more and more maximally mixed in the large squeezing limit such that $C_E/S_E\sim z^{-1}\rightarrow 0$. As we have mentioned before this means that the corresponding reduced density matrix can be approximated as proportional to the identity operator to the extent that its Renyi entropies are independent of $n$ and thus we have a flat entanglement spectra. Remarkably, in the holographic context one can produce the $n$-independent Renyi entropies by considering specific semiclassical states, the so-called fixed-area states \cite{Dong:2018seb,Akers:2018fow,Botta-Cantcheff:2020ywu,Arias:2021ilh}.\footnote{We thank Mohammad Hasan Vahidinia for bringing this to our attention and raising the following question.} It is an interesting question whether or not a more concrete connection can be found between the squeezing states in the field theory and fixed-area states
of quantum gravity.

Of course, eq. \eqref{largezCE1} is consistent with the Page’s argument that in an arbitrary quantum state the entropy is close to maximal. Hence we expect entanglement entropy to scale with the volume (instead of the area) of the entangling region. Our results show that this interesting behavior also holds for the capacity of entanglement. Moreover, the measures are both sensitive to the UV cutoff, but the ratio is finite and thus at least for the free massless scalar field theory, the ratio of the leading terms is scheme independent. We found similar results for other entanglement measures which are consistent with the asymptotic expansions in the large squeezing limit.

To close our discussion, we would like to recall that our result for the capacity of entanglement for a family of random  states, where $\overline{C_E}$ is continuous, contrasts with the holographic computations correspond to the black hole evaporation process reported in \cite{Kawabata:2021hac,Kawabata:2021vyo}. Indeed, we expect that a meaningful comparison between the field theory and holographic results is achieved by considering strongly coupled theories with a large number of degrees of freedom. This investigation is beyond the scope of the present paper. It would be interesting to explore the behavior of $C_E$ in general setups including rational and holographic CFTs \cite{Hartman:2013mia,Hartman:2014oaa,Caputa:2014vaa,Asplund:2015eha}. Some additional topics to explore include generalizing our results to higher dimensions \cite{Katsinis:2024sek} or to non-relativistic models \cite{Ardonne:2003wa,MohammadiMozaffar:2017nri,He:2017wla,Boudreault:2021pgj}. Another interesting direction is to study the scaling of capacity of entanglement in theories which exhibit a volume law scaling of the entanglement entropy \cite{Shiba:2013jja,Karczmarek:2013xxa,Mollabashi:2014qfa,MohammadiMozaffar:2015clv}. In addition, we have been focused on the exploration of the capacity only in a free scalar theory, but it should be feasible to extend our work to more realistic cases with non-trivial interactions where we believe the capacity will exhibit more interesting features \cite{Cotler:2015zda}. We leave the details of
some interesting problems for future study \cite{MohammadiMozaffar:2024uiy}.

\subsection*{Acknowledgements}

The author would like to thank Ali Mollabashi and Mohammad Hasan Vahidinia for reading the
draft and fruitful discussions. We are also very grateful to Dimitrios Katsinis, Georgios Pastras and Nikolaos Tetradis for clarifications about aspects of ref. \cite{Katsinis:2023hqn}. This work is based upon research funded by Iran National Science Foundation (INSF) under project No. 4013637.

%

\end{document}